\documentclass[a4paper]{article}

\usepackage{INTERSPEECH2020}
\usepackage{float}
\usepackage{amsmath,graphicx,amssymb}
\usepackage{verbatim}
\usepackage[dvipsnames]{xcolor}
\usepackage{tikz}
\usetikzlibrary{decorations.text}
\usetikzlibrary{calc}

\makeatletter
\newcommand\notsotiny{\@setfontsize\notsotiny\@vipt\@viipt}
\newcommand{\class}[1]{\small \texttt{#1}}

\title{An open-source voice type classifier for child-centered daylong recordings}
\name{Marvin Lavechin$^{1,2}$, Ruben Bousbib$^{1,2}$, Herv\'e Bredin$^3$, Emmanuel Dupoux$^{1,2}$, Alejandrina Cristia$^{1}$}

\address{
  $^1$ENS-PSL/CNRS/EHESS, Paris, France;\\ 
  $^2$INRIA Paris, France; 
  \\
  $^3$LIMSI, CNRS, Univ. Paris-Sud, Université Paris-Saclay, Orsay, France.
  \thanks{This work was performed using HPC resources from GENCI-IDRIS (Grant 2020-A0071011046). It also benefited from the support of ANR-16-DATA-0004 ACLEW (Analyzing Child Language Experiences collaborative project), ANR-17-CE28-0007 (LangAge), ANR-14-CE30-0003 (MechELex), ANR-17- EURE-0017 (Frontcog), ANR-10-IDEX-0001-02 (PSL), ANR-19-P3IA-0001 (PRAIRIE 3IA Institute), and the J. S. McDonnell Foundation Understanding Human Cognition Scholar Award.}}
\email{$^{}$marvinlavechin@gmail.com, 
alecristia@gmail.com}

\begin{document}

\maketitle
\begin{abstract}
Spontaneous conversations in real-world settings such as those found in child-centered recordings have been shown to be amongst the most challenging audio files to process. Nevertheless, building speech processing models handling such a wide variety of conditions would be particularly useful for language acquisition studies in which researchers are interested in the quantity and quality of the speech that children hear and produce, as well as for early diagnosis and measuring effects of remediation. In this paper, we present our approach to designing an open-source neural network to classify audio segments into vocalizations produced by the child wearing the recording device, vocalizations produced by other children, adult male speech, and adult female speech. To this end, we gathered diverse child-centered corpora which sums up to a total of 260 hours of recordings and covers 10 languages. Our model can be used as input for downstream tasks such as estimating the number of words produced by adult speakers, or the number of linguistic units produced by children. Our architecture combines SincNet filters with a stack of recurrent layers and outperforms by a large margin the state-of-the-art system, the Language ENvironment Analysis (LENA) that has been used in numerous child language studies.
\end{abstract}


\section{Introduction and related work}

In the past, language acquisition researchers' main material was short recordings \cite{hart1995} or times of in-person observations \cite{wells1979}. However, investigating the language phenomenon in this manner can lead to biased observations, potentially resulting in divergent conclusions \cite{bergelson2019day}. More recently, technology has allowed researchers to efficiently collect and analyze recordings over a whole day. By the combined use of a small wearable device and speech processing algorithms, one can get meaningful insights of children's daily language experiences. While daylong recordings are becoming a central tool for studying how children learn language, a relatively small effort has been made to propose robust and bias-free speech processing models to analyze such data. It may however be noticed that some collaborative works that benefit both the speech processing and the child language acquisition communities have been done. In particular, we may cite Homebank, an online repository of daylong child-centered audio recordings \cite{homebank} that allow researchers to share data more easily. Some efforts have also been made to gather state-of-the-art pretrained speech processing models in DiViMe \cite{divime}, a user-friendly and open-source virtual machine. Challenges and workshops using child-centered recordings \cite{ryant2019,garcia2019}, also attracted the attention of the speech processing community. Additionally, the task of classifying audio events has often been addressed in the speech technology literature. In particular, the speech activity detection task \cite{ryant2013speech} or the acoustic event detection problem \cite{acousticmedia1, acoustic_spectro} are similar to the voice type classification task we address in this paper.

Given the lack of open-source and easy-to-use speech processing models for treating child-centered recordings, researchers have been relying, for the most part, on the Language ENvironment Analysis (LENA) software \cite{xu_lenatm_2009} to extract meaningful information about children's language environment. This system will be introduced in more detail in the next section.

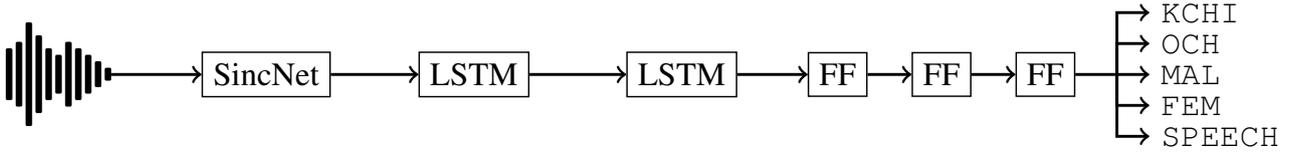
\begin{figure*}[h]
\tikzstyle{element}=[rectangle,draw, font=\small]
\tikzstyle{arrow}=[->,thick]
\tikzstyle{label}=[right, font=\footnotesize]

\centering
\resizebox{\textwidth}{!}{%
    \begin{tikzpicture}
        \node[inner sep=0pt,text width=10mm] (audiotop) at (-1.9,0.9) {};
        \node[inner sep=0pt,text width=10mm] (audiobot) at (-1.9,-0
    .6) {};
        \node[inner sep=0pt,text width=10mm] (audio) at (-2,0)
                {\includegraphics[width=10mm]{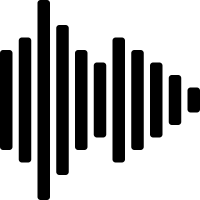}};
        \node[label] (KCHI) at (8.5, 0.60) {\class{KCHI}};
        \node[label] (OCH) at (8.5,0.30) {\class{OCH}};
        \node[label] (MAL) at (8.5,0.0) {\class{MAL}};
        \node[label] (FEM) at (8.5,-0.30) {\class{FEM}};
        \node[label] (SPEECH) at (8.5,-0.60) {\class{SPEECH}};

        \node[element] (sincnet) at (0,0) {SincNet};
        \node[element] (lstm1) at (2,0) {LSTM};
        \node[element] (lstm2) at (4,0) {LSTM};
        \node[element] (ff1) at (5.5,0) {FF};
        \node[element] (ff2) at (6.5,0) {FF};
        \node[element] (ff3) at (7.5,0) {FF};
      
        \draw[arrow] (audio) -- (sincnet);
        \draw[arrow] (sincnet) -- (lstm1) coordinate[midway] (aux){};
        \draw[arrow] (lstm1) -- (lstm2);
        \draw[arrow] (lstm2) -- (ff1);
        \draw[arrow] (ff1) -- (ff2);
        \draw[arrow] (ff2) -- (ff3);

        \draw[thick] (ff3.east) -- ([xshift=0.4cm]ff3.east);
        \draw[arrow] ([xshift=0.4cm]ff3.east) |- (KCHI.west);
        \draw[arrow] ([xshift=0.4cm]ff3.east) |- (OCH.west);
        \draw[arrow] ([xshift=0.4cm]ff3.east) |- (MAL.west);
        \draw[arrow] ([xshift=0.4cm]ff3.east) |- (FEM.west);
        \draw[arrow] ([xshift=0.4cm]ff3.east) |- (SPEECH.west);
    \end{tikzpicture}
}
\caption{Proposed architecture. The network takes the raw waveform of a 2s audio chunk as input and passes on to SincNet \protect\cite{sincnet}. The low-level representations learnt by SincNet are then fed to a stack of two bi-directional LSTMs, followed by three feed-forward layers. The output layer is activated by a sigmoid function that returns a score ranging between 0 and 1 for each of the classes.}
\label{fig:architecture}
\end{figure*}
\subsection{The LENA system}

The LENA system consists of a small wearable device combined with an automated vocal analysis pipeline that can be used to study child language acquisition. The audio recorder has been designed to be worn by young children as they go through a typical day. In the current LENA system, after a full day of audio has been been captured by the recorder, the audio files are transferred to a cloud and analyzed by signal processing models. These latter have been trained on 150 hours of proprietary audio collected from recorders worn by American English-speaking children. The speech processing pipeline consists of the following steps \cite{xu_lenatm_2009, gilkerson_transcriptions,ganek_language}:

\begin{enumerate}
    \item[1] First, the audio is segmented into mutually exclusive categories that include: key child vocalizations (i.e., vocalizations produced by the child wearing the recording device), adult male speech, adult female speech, other child speech, overlapping sounds, noise, and electronic sounds.
    \item[2] The key child vocalization segments are further categorized into speech and non-speech sounds. Speech encompasses not only words, but also babbling and pre-speech communicative sounds (such as squeals and growls). Child non-speech sounds include emotional reactions such as cries, screams, laughs and vegetative sounds such as breathing and burping.
    \item[3] A model based on a phone decoder estimates the number of words in each adult speech segment.
    \item[4] Further analyses are performed to detect conversational turns, or back and forth alternations between the key child and an adult.
\end{enumerate}

The LENA system has been used in multiple studies covering a wide range of expertise including a vocal analysis of children suffering from hearing loss \cite{hearing_LENA_study}, the assessment of a parent coaching intervention \cite{parent_coaching_LENA}, and a study of autism spectrum disorders \cite{autism_LENA}. An extensive effort has been made to assess the performance of the LENA speech processing pipeline \cite{lena_eval1,lena_eval2,lena_eval3}.

Despite its wide use in the child language community, LENA imposes several limiting factors to scientific progress. First, as their software is closed source, there is no way to build upon their models to improve  performance, and we cannot be certain about all design choices and their potential impact on  performance. Moreover, since their models have been trained only on American English-speaking children recorded with one specific piece of hardware in urban settings, the model might potentially be overfit to these settings, with a loss of generalization to other languages, cultures, and recording devices.


\subsection{The present work}

Our work aims at proposing a viable open-source alternative to LENA for classifying audio frames into segments of key child vocalizations, adult male speech, adult female speech, other child vocalizations, and silence. The general architecture is presented in \ref{e2e}. Additionally, we gathered multiple child-centered corpora covering a wide range of conditions to train our model and compare it against LENA. This data set is described in further details in \ref{data}. 

\section{Experiments}

\subsection{End-to-end voice type classification}\label{e2e}

The voice type classification problem can be described as the task of identifying voice signal sources in a given audio stream. It can be tackled as a multi-label classification problem where the input is the audio stream divided into $N$ frames $\boldsymbol{S} = \{s_1, s_2, \ldots, s_N\}$ and the expected output is the corresponding sequence of labels $\boldsymbol{y} = \{\boldsymbol{y_1},\boldsymbol{y_2},\ldots,\boldsymbol{y_N}\}$ where each $\boldsymbol{y_i}$ is of dimension $K$ (the number of labels) with $y_{i,j} = 1$ if the $j^{th}$ class is activated, $y_{i,j} = 0$ otherwise. Note that, in the multi-label setup, multiple classes can be activated at the same time.

At training time, fixed-length sub-sequences made of multiple successive frames, are drawn randomly from the training set to form mini-batches of size $M$.

As illustrated in Figure~\ref{fig:architecture}, these fixed-length sub-sequences are processed by a SincNet \cite{sincnet} that aims at learning meaningful filter banks specifically customized to solve the voice type classification task. These low-level signal representations are then fed into a stack of bi-directional long short-term memory (LSTM) layers followed by a stack of feed-forward (FF) layers. Finally, the sigmoid activation function is applied to the final output layer of dimension $K$ so that each predicted score $\hat{y}_{i,j}$ consists of a number ranging between 0 and 1.
The network is trained to minimize the binary cross-entropy loss:

\begin{equation}
    \mathcal{L} = - \dfrac{1}{KM} \sum_{i=1}^M \sum_{j=1}^K y_{i,j} \log \hat{y}_{i,j} + \left(1-y_{i,j} \right) \log \left(1 - \hat{y}_{i,j} \right)
    \label{loss}
\end{equation} 

At test time, audio files are processed using overlapping sliding sub-sequences of the same length as the one used in training. For each time step $t$, and each class $j$, this results in several overlapping sequences of prediction scores, which are averaged to obtain the final score for class $j$. Finally, time steps with prediction scores greater than a tunable threshold $\sigma_j$ are marked as being activated for the class $j$.

Our use case considers $K = 5$ different classes or sources which are: 
\begin{itemize}
    \item \class{KCHI}, for key-child vocalizations, i.e., vocalizations produced by the child wearing the recording device
    \item \class{OCH}, for all the vocalizations produced by other children in the environment
    \item \class{FEM}, for adult female speech
    \item \class{MAL}, for adult male speech
    \item \class{SPEECH}, for when there is speech
\end{itemize}

As the LENA voice type classification model is often used to sample audio in order to extract segments containing the most speech, it appeared to us that it was useful to consider a class for speech segments produced by any type of speaker. Moreover, in our data set, some of the segments have been annotated as \class{UNK} (for unknown) when the annotator was not certain of which type of speaker was speaking (See Table \ref{tab:bbt}). Considering the \class{SPEECH} class allows our model to handle these cases.

One major design difference with the LENA model is that we chose to treat the problem as a multi-label classification task, hence multiple classes can be activated at the same time (e.g., in case of overlapping speech). In contrast, LENA treats the problem as a multi-class classification task where only one class can be activated at a given time step. In the case of overlapping speech, LENA model returns the \class{OVL} class (which is also used for overlap between speech and noise). More details about the performance obtained by LENA on this class can be found in \cite{lena_eval1}. 

\subsection{Datasets}\label{data}

\begin{table*}[h]
    \centering
    \caption{Description of the BabyTrain data set. Child-centered corpora included cover a wide range of conditions (including different languages and recording devices). ACLEW-Random is kept as a hold-out data set on which LENA and our model are compared. DB correpond to datasets that can be found on Databrary, HB the ones that can be found on Homebank.}
    \label{tab:bbt}
    \resizebox{\textwidth}{!}{
    \begin{tabular}{llllrrrrrr}
        \toprule
        & & & & & \multicolumn{5}{c}{Cumulated utterance duration} \\
        \cmidrule{6-10}
        Corpus & Access & LENA-recorded? & Language & Tot. Dur. & \class{KCHI} & \class{OCH} & \class{MAL} & \class{FEM} & \class{UNK}\\
        \midrule
        \multicolumn{9}{c}{\textbf{BabyTrain}} \\
        \midrule
        ACLEW-Starter & mixture (DB) & mostly & Mixture & 1h30m & 10m & 5m & 6m & 20m & 0m\\
        Lena Lyon & private (HB) & yes & French & 26h51m & 4h33m & 1h14m & 1h9m & 5h02m & 1h0m\\
        Namibia & upon agreement & no & Ju$\vert$'hoan & 23h44m & 1h56m & 1h32m & 41m & 2h22m & 1h01m\\
        Paido & public (HB) & no & Greek, Eng., Jap. & 40h08m & 10h56m & 0m & 0m & 0m & 0m\\
        Tsay & public (HB) & no & Mandarin & 132h02m & 34h07m & 2h08m & 10m & 57h31m & 28m\\
        Tsimane & upon agreement & mostly & Tsimane & 9h30m & 37m & 23m & 11m & 28m & 0m\\
        Vanuatu & upon agreement & no & Mixture & 2h29m & 12m & 5m & 5m & 9m & 1m\\
        WAR2 & public (DB) & yes & English (US) & 50m & 14m & 0m & 0m & 0m & 9m\\
        \midrule
        \multicolumn{9}{c}{\textbf{Hold-out set}} \\
        \midrule
        ACLEW-Random & private (DB) & yes & Mixture & 20h & 1h39m & 45m & 43m & 2h48m & 0m\\
        \bottomrule
    \end{tabular}
    }
\end{table*}

In order to train our model, we gathered multiple child-centered corpora data \cite{data1, data2, data3, data4, data5, data6, data7, data8, data9, data10} drawn from various child-centered sources, several of which were not daylong. Importantly, the recordings used for this work cover a wide range of environments, conditions and languages and have been collected and annotated by numerous field researchers.

We will  refer to this data set as BabyTrain, of which a broad description is given in Table \ref{tab:bbt}. 

We split the BabyTrain data set into a training, development and test sets, containing approximately 60\%, 20\% and 20\% of the audio duration respectively. We applied this split such that files associated to a given key child were included in only one of the three sets, splitting children up within each of the 8 corpora of BabyTrain. The only exception was WAR2, too small to be divided, and therefore put in the training set in its entirety.

In order to ensure that our models generalize well enough to unseen data, and to compare the performance with the LENA system, we kept the ACLEW-Random as a hold-out data set.

\subsection{Evaluation metric}

For each class, we use the F-measure between precision and recall, such as implemented in {\small \texttt{pyannote.metrics}}~\cite{pyannote.metrics} to evaluate our systems: 
\begin{equation*}
\text{F-measure} = \dfrac{2 \times \text{precision} \times \text{recall}}{\text{precision} \times \text{recall}}
\end{equation*}
\noindent where $\text{precision} = \text{tp} / (\text{tp} + \text{fp})$ and $\text{recall} = \text{tp} / (\text{tp} + \text{fn})$ with:
\begin{itemize}
    \item $\text{tp}$ the duration of true positives
    \item $\text{fp}$ the duration of false positives
    \item $\text{fn}$ the duration of false negatives
\end{itemize}

\noindent We select our models by averaging the F-measure across the 5 classes. Note that these 5 metrics have been computed in a binary fashion, where the predictions of our model for a given class were compared to all reference speaker turns such as provided by the human annotations (no matter if the latter were overlapping or not). In diarization studies, the choice of a collar around every reference speaker turns is often made to account for inaccuracies in the reference labels. We chose not to do so, consequently all numbers reported in this paper can be considered as having a collar equal to 0.

\subsection{Implementation details}

Figure \ref{fig:architecture} illustrates the broad architecture used in all experiments. For SincNet, we use the configuration proposed by the authors of the original paper \cite{sincnet}. All LSTMs and inner feed-forward layers  have a size of 128 and use \emph{tanh} activations. The last feed-forward layer uses a sigmoid activation function.

Data augmentation is applied directly on the waveform using additive noise extracted from the MUSAN database \cite{musan} with a random target signal-to-noise ratio ranging from 5 to 20 dB.
The learning rate is set up by a cyclical scheduler \cite{cyclical}, each cycle lasting for 1.5 epoch.

Since we address the problem in a multi-label classification fashion, multiple classes can be activated at the same time. For the reference turns, the \class{SPEECH} class was considered to be activated whenever one (or more) of the \class{KCHI}, \class{CHI}, \class{FEM}, \class{MAL} or \class{UNK} class was activated. The \class{UNK} class (see Table \ref{tab:bbt}) corresponds to cases when the human annotator could hear that the audio contained speech or vocalizations, without being able to identify the voice source. This class does contribute in activating the \class{SPEECH} class, but our model does not return a score for it.

\subsection{Evaluation protocol}

For all experiments, the neural network is trained for 10 epochs (approximately 2400 hours of audio) on the training set. The development set is used to choose the actual epoch and thresholds $\{\sigma_j\}_{j=1}^K$ that maximizes the average F-measure between precision and recall across classes.

We report both the in-domain performance (computed on the test set of BabyTrain) and the out-of-domain performance (computed on the hold-out set, ACLEW-Random). We compare our model with the LENA system on the hold-out set.

\begin{table*}[h]
    \centering
    \caption{In-domain performance in terms of F-measure between precision and recall. The "\textit{Ave.}" column represents the F-measure averaged across the 5 classes. Numbers are computed on the test set from which the Paido corpora has been removed. Performance on the development set are reported using small font size. We report two variants, the first one is based on 5 binary models trained separately on each of the class, the second one consists of a single model trained in a multitask fashion}
    \label{tab:in_perf}
    \begin{tabular}{rrcccccc}
        \toprule
         Train/Dev. & System & \class{KCHI} & \class{OCH} & \class{MAL} & \class{FEM} & \class{SPEECH} & Ave.\\
        \midrule
        without Paido & binary & 76.1 {\scriptsize{79.2}} & 22.5 {\scriptsize{28.7}} & 37.8 {\scriptsize{38.9}} & 80.2 {\scriptsize{83.5}} & 88.0 {\scriptsize{89.3}} & 60.9 {\scriptsize{63.9}}  \\
        with Paido & multi & 75.8 {\scriptsize{78.7}} & 25.4 {\scriptsize{30.3}} & 40.1 {\scriptsize{43.2}} & 82.3 {\scriptsize{83.9}} & 88.2 {\scriptsize{90.1}} & 62.3 {\scriptsize{65.2}} \\
        without Paido & multi & 77.3 {\scriptsize{80.6}} & 25.6 {\scriptsize{30.6}} & 42.2 {\scriptsize{43.7}} & 82.4 {\scriptsize{84.2}} & 88.4 {\scriptsize{90.3}} & 63.2 {\scriptsize{65.9}} \\
        \bottomrule
    \end{tabular}
\end{table*}

\section{Results}

We evaluate two different approaches, one consisting of 5 models trained separately for each of the class (referred as binary), and one consisting of a single model trained jointly on all the classes (referred as multitask). At first, both in the binary and the multitask scenario, architectures shared the same set of hyper-parameters. Only the dimension of the output layer differed. Results indicated that multitask approaches were significantly better than binary ones, which seems to show that sharing weights during training helps better learn the boundaries between the different classes.
 
To further improve the performance of our model, we tried multiple sets of hyper-parameters (varying the number of filters, the number of LSTM and FF layers, and their size). However, no significant differences have been observed among the different architectures. The retained architecture consists of 256 filters of length L = 251 samples, 3 LSTM layers of size 128, and 2 FF layers of size 128.

Finally, removing Paido from the training and development set led to improvements on the other test domains, as well as the hold-out set, while the performance on the Paido domain remained high. Indeed, we observed a F-measure of 99 on the \class{KCHI} class for the model trained with Paido as compared to 89 for the model trained without it. This difference can be explained by a higher amount of false alarms returned by the model trained without it. The Paido domain is quite far from our target domain since it consists of laboratory recordings of words in isolation spoken by children, and thus it is reasonable to think that removing it leads to better models. 

\subsection{In-domain performance}

Since LENA can only be evaluated in data collected exclusively with the LENA recording device and BabyTrain contains a mixture of devices, we do not report on LENA in-domain performance.  Additionally, comparing performance on a domain that would have been seen during the training by our model but not by LENA would have unfairly advantaged us.

Table \ref{tab:in_perf} shows results in terms of F-measure between precision and recall on the test set for each of the 5 classes. The best performance is obtained for the \class{KCHI}, \class{FEM}, and \class{SPEECH} classes, which correspond to the 3 classes that are the most present in BabyTrain (See Table \ref{tab:bbt}). Performance is lower for the \class{OCH} class and \class{MAL} classes, with an F-measure of 25.6 and 42.2 respectively, most likely due to the fact that these two classes are underrepresented in our data set. The F-measure is lowest for the \class{OCH} class. In addition to being underrepresented in the training set, utterances belonging to the \class{OCH} class can easily be confused with \class{KCHI} utterances since the main feature that differentiates these two classes is the average distance to the microphone.

The multitask model consistently outperforms binary ones. When training in a multitask fashion, increases are higher for the lesser represented classes, namely \class{OCH} and \class{MAL}. Additionally, removing Paido leads to an improvement of 0.9 in terms of average F-measure on the other domains. 

\subsection{Performance on the hold-out data set}

\begin{table}[htb]
    \centering
    
    \caption{Performance on the hold-out data set in terms of F-measure between precision and recall. "\textit{Ave.}" column represents the F-measure averaged across the 5 classes. The hold-out data set has never been seen during the training, neither by LENA, nor by our model.}
    \label{tab:out_perf}
    \resizebox{\columnwidth}{!}{
    \begin{tabular}{rrcccccc}
        \toprule
        Train/Dev. &System & \class{KCHI} & \class{OCH} & \class{MAL} & \class{FEM} & \class{SPEECH} & Ave.\\
        \midrule
        english (USA) & LENA & 54.9 & 28.5 & 37.2 & 42.6 & 70.2 & 46.7 \\
        without Paido & binary & 67.6 & 23.0 & 31.6 & 62.6 & 77.6 & 52.5\\
        with Paido & multi & 66.4 & 19.9 & 39.9 & 63.0 & 77.6 & 53.3\\
        without Paido & multi & \textbf{68.7} & \textbf{33.2} & \textbf{42.9} & \textbf{63.4} & \textbf{78.4} & \textbf{57.3}\\
        \bottomrule
    \end{tabular}
    }
\end{table}

Table \ref{tab:out_perf} shows performance of LENA, our binary variant, and our multitask variant on the hold-out data set. As observed on the test set, the model trained in a multi-task fashion shows better performance than the models trained in a binary fashion. Removing Paido leads to a performance increase of 4 points on the average F-measure.


Turning to the comparison with LENA, both the LENA model and our model show lower performance for the rarer \class{OCH} and \class{MAL} classes. 
Our model outperforms the LENA model by a large margin.  We observe an absolute improvement in terms of F-measure of  13.8 on the \class{KCHI} class, 4.6 on the \class{OCH} class, 5.6 on the \class{MAL} class, 20.8 on the \class{FEM} class, and 8.1 on the \class{SPEECH} class. This leads to an absolute improvement of 10.6 in terms of F-measure averaged across the 5 classes.

\section{Reproducible research}

All the code has been implemented using \small\texttt{pyannote.audio} \cite{pyannote.audio}, a python open-source toolkit for speaker diarization. Our own code, easy-to-use scripts to apply the pretrained model can be found on our GitHub repository \footnote{\texttt{https://github.com/MarvinLvn/voice-type-classifier}}, which also includes confusion matrices and a more extensive comparison with LENA. As soon as required agreements will be obtained, we plan to facilitate access to the data by hosting them on Homebank.

\section{Conclusion}

In this paper, we gathered recordings drawn from diverse child-centered corpora that are known to be amongst the most challenging audio files to process, and proposed an open-source speech processing model that classifies audio segments into key child vocalizations, other children vocalizations, adult male speech, and adult female speech. We compared our approach with a homologous system, the LENA software, which has been used in numerous child language studies. Our model outperforms LENA by a large margin and will, we hope, lead to more accurate observations of early linguistic environments. Our work is part of an effort to strengthen collaborations between the speech processing and the child language acquisition communities. The latter have provided data as that used here, as well as interesting challenges \cite{ryant2019,garcia2019}. Our paper is an example of  the speech processing community returning the favor by providing robust models that can handle spontaneous conversations in real-world settings. 

\clearpage
\bibliographystyle{IEEEtran}
\bibliography{refs}

\end{document}